\documentclass[12pt,a4paper]{article}
\usepackage{amsmath}
\usepackage[dvips]{graphicx}
\input epsf
\usepackage{times}
\begin{document}
\begin{titlepage}
\title{Cosmic rays and antishadowing}
\author{S. M. Troshin\footnote{email: Sergey.Troshin@ihep.ru},
 N. E. Tyurin\\[1ex]
\small  \it Institute for High Energy Physics,\\
\small  \it Protvino, Moscow Region, 142281, Russia}
\normalsize
\date{}
\maketitle

\begin{abstract}
We note that antishadowing could help in the explanation  of cosmic
rays regularities such as knee in the energetic spectrum and existence of
penetrating and long-flying particles.\\[2ex]
\end{abstract}
\end{titlepage}
\setcounter{page}{2}

Cosmic rays are the charged nuclei arriving from the outside of the solar system.
The cosmic rays investigations are an important source of astrophysical information
(cf. e.g. \cite{der})
and simultaneously they provide a window to the future of accelerator
studies\footnote{It should be noted however that the results for the total
cross--section extracted from cosmic rays measurements significantly rely
on particular  model, because cosmic rays do not provide information on elastic
scattering cross--section \cite{engel}}.

\begin{figure}[thb]
\begin{center}
\includegraphics[width=60mm]{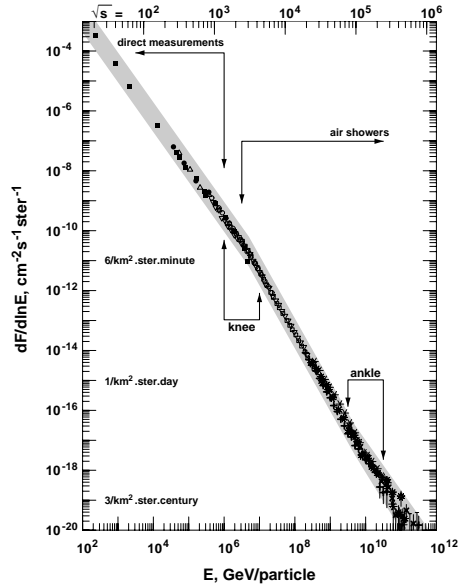}
\end{center}
\small{\caption{Energy spectrum of the cosmic rays, figure from the
 first reference of  \cite{crrev}.}}
\label{ts:fig1}
\end{figure}
Investigations of cosmic rays give us a clue that the hadron
interaction and mechanism of particle
generation is changing in the region of $\sqrt{s}=3-6$ GeV\cite{crrev,her}. Indeed, the
energy spectrum which follows simple power-like law $F(E)=cE^{-\gamma}$ changes
its slope in this energy region and becomes steeper: index $\gamma$ increases from
$2.7$ to $3.1$. It is important that the knee in the energy spectrum appears in the same
energy region where the penetrating and long--flying particles also start to appear in
the extended air showers (EAS):
the absorbtion length is also changing from $\lambda=90$ $g/cm^2$ to
$\lambda=150$ $g/cm^2$ (cf. \cite{crrev}). There is also specific feature of the
events at the energies beyond knee such as alignment.
The above phenomena  were interpreted as a result of appearance among the secondaries
 of the new particles which have a small inelastic cross--section and/or
small inelasticity. These new particles can be associated with a manifestation of the
supersymmetry, quark--gluon plasma formation and other new mechanisms. However, there
is another possibility to treat those cosmic rays phenomena as the avatars of the antishadow
 scattering mode
at such energies \cite{ashd}.

Unitarity of the scattering matrix $SS^+=1$ implies, in principle, an
existence at high energies $s>s_0$, where $s_0$ is a threshold\footnote{Model
estimates show that
antishadow scattering mode starts to develop right beyond Tevatron energies, i.e.
$\sqrt{s_0}\simeq 2$ TeV\cite{s0}}
 of the new scattering mode --
antishadow one. It has been revealed in \cite{ashd} and
 described in some detail (cf.
\cite{echn} and references therein) and the most important feature
of this mode is the self-damping of the  contribution from the
inelastic channels.

Antishadowing  leads to
 asymptotically dominating role of elastic scattering.
The cross--section of inelastic processes rises with energy as $\ln s$, while
  elastic and total cross--sections behave asymptotically as $\ln^2 s$.
The antishadow scattering mode could be definitely
revealed at  the LHC energies and the phenomena observed in the
cosmic rays studies confirm it.
Starting at some threshold energy $s_0$ (where amplitude reaches the black disk
 limit at $b=0$), antishadowing can occur at higher energis
 in the limited region of impact parameters $b<R(s)$ (while
 at large impact parameters only shadow scattering mode can be
 realized). Note that a shadow scattering mode can exist without antishadowing,
 but the opposite is not true.
\begin{center}
\begin{figure}[hbt]
\hspace*{2cm}\epsfxsize= 100  mm  \epsfbox{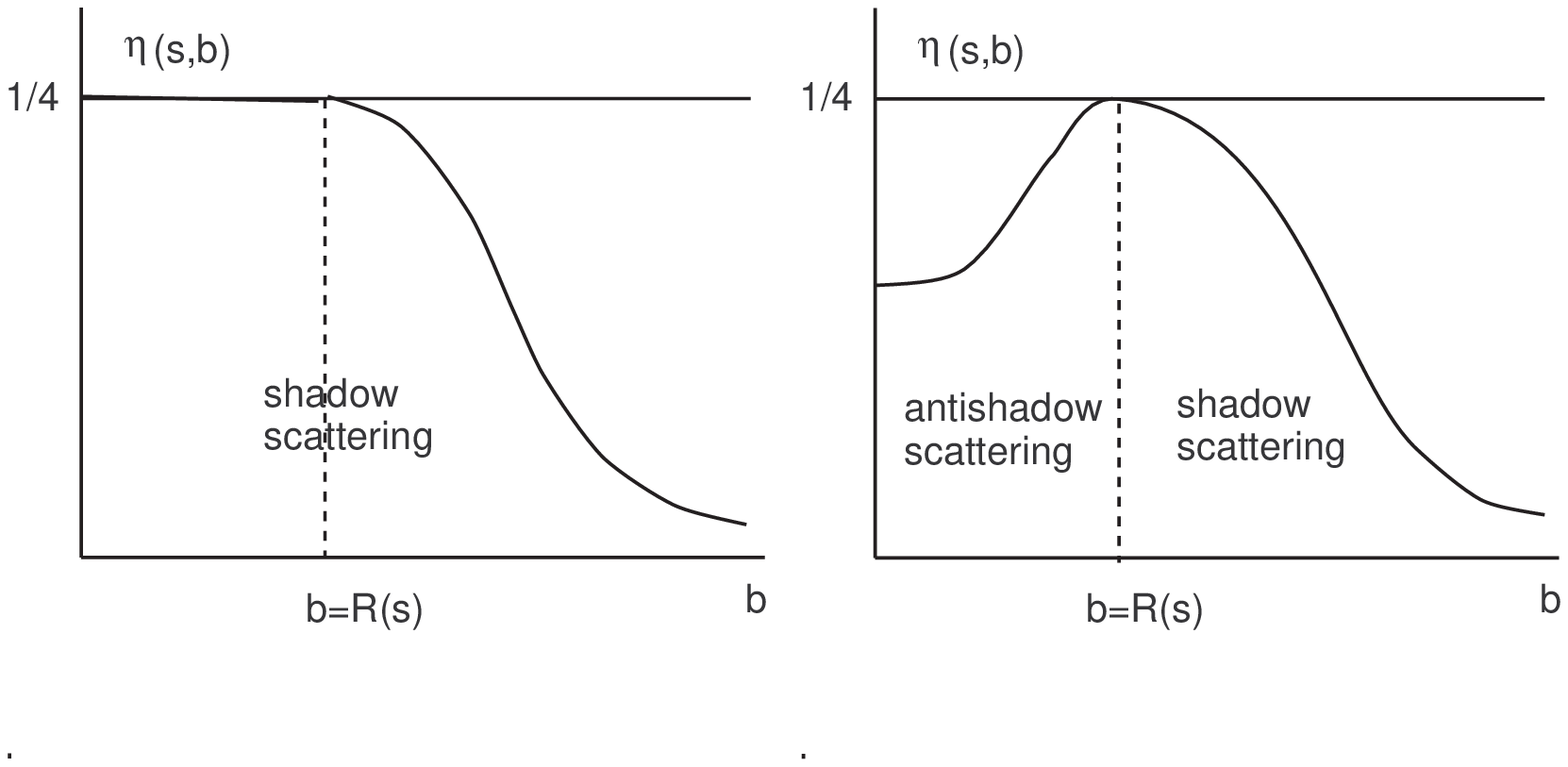}
{\small\caption{Impact parameter dependence of the inelastic overlap
function
 in the standard unitarization scheme (left panel) and in the unitarization scheme
 with antishadowing (right panel).}}
 \end{figure}
\end{center}
The inelastic overlap function $\eta(s,b)$ becomes peripheral
 when energy is beyond $s=s_0$.
At such energies the inelastic overlap function reaches its maximum
 value at $b=R(s)$ where $R(s)$ is the interaction radius.
So, beyond the transition energy range there are two regions in impact
 parameter space: the central region
of antishadow scattering at $b< R(s)$ and the peripheral region of
shadow scattering at $b> R(s)$. The impact parameter
dependence of the inelastic channel contribution
$\eta(s,b)$ at $s>s_0$ are represented in Fig. 2 for the case of
standard unitarization scheme  and for the unitarization scheme with
anishadowing.

At the energies ${s}>{s_0}$ small impact parameter scattering
is mostly elastic one. Thus head--on colliding particles  will provide appearance
of penetrating long-flying component in the  EAS
 and such particles will  spend only
small part  of
their energy for the production of secondaries. The head-on collisions
will lead to smaller  number of secondary particle and it will provide
 faster decrease of the energy spectrum of cosmic rays, i.e. it
  will result in the appearance of the knee. This qualitative picture
  will be explained in more detail  in what follows.

Antishadowing leads to suppression of particle
production at small impact parameters, and the main contribution to
the integral  multiplicity $\bar n(s)$
\begin{equation}\label{mm}
\bar n(s)= \frac{\int_0^\infty  \bar n
(s,b)\eta(s,b)bdb}{\int_0^\infty \eta(s,b)bdb}
\end{equation}
comes from the region of $b\sim R(s)$.

Due to peripheral form of the inelastic overlap function the
secondary particles will be mainly
 produced at impact parameters $b\sim R(s)$ and this could lead to the events with
 alignment observed in cosmic rays  and also to the imbalance
 between orbital angular momentum in the initial and final states since particles
 in the final state will carry  out large orbital
 angular momentum.
To compensate this orbital momentum spins of secondary particles should  become
  lined up, i.e. the spins of the produced particles should demonstrate
   significant
  correlations when the antishadow scattering mode appears \cite{sc}. Thus, the
   observed phenomena of
  alignment in cosmic rays events
  and predicted spin correlations of final particles should have a common origin.

Antishadowing  leads to the nonmonotonous energy dependence of gap survival
probability \cite{gsp}. The gap survival probability, namely the probability to keep away
inelastic interactions  which can result in filling up by hadrons
the large rapidity gaps, reaches its
minimal values at the Tevatron highest energy and this is due to the fact that
 the scattering at this energy is very close to
the black disk limit at $b=0$ (Fig. 3).
\begin{figure}[thb]
\begin{center}
\includegraphics[width=60mm]{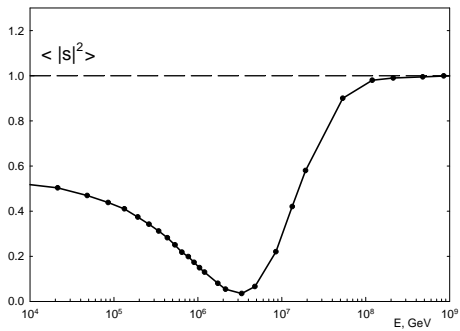}
\end{center}
\small{\caption{Energy dependence of gap survival probability. }}
\label{fig3}
\end{figure}
 It is clear that its higher value
means higher fraction for diffractive  component and consequently
the increasing of this component would result  in the enhancement of
the relative fraction
of protons in  the observed cosmic rays spectrum. Otherwise, decreasing of this quantity
will lead to increase of pionization  component and consequently to the increasing
number of muons
  observed as  multi-muon events.
Experiment reveals that relative fraction of protons in cosmic rays also shows
nonmonotonous energy dependence \cite{hern}. To explain such dependence an additional
component is introduced {\it ad hoc} at the energies above $3\cdot 10^7$ GeV. It was shown
that account
of the antishadowing makes an introduction of this {\it ad hoc} component unnecessary.

The  inelasticity parameter $K$,
which is defined as ratio of the energy going to inelastic processes to the total energy, is
 important for the interpretation
of the cosmic rays cascades developments. Its energy dependence is not
 clear and number of models predict the decreasing energy dependence while other
 models insist on the increasing energy behaviour at high energies
 (cf. e.g. \cite{shabel}). Adopting simple ansatz of geometrical models where parameter
 of inelasticity
 is related to inelastic overlap function we can use the following equation for
 $\langle K \rangle$ \cite{dias}
 \[
 \langle K \rangle=4\frac{\sigma_{el}}{\sigma_{tot}}
 \left(1-\frac{\sigma_{el}}{\sigma_{tot}}\right)
\]
to get a qualitative knowledge on the inelasticity energy dependence.
The estimation of inelasticity  based on the particular model with
 antishadowing \cite{s0} leads to increasing dependence of inelasticity with energy
 till $E\simeq 4\cdot 10^7$ GeV. In this region inelasticity reaches maximum value
 $\langle K \rangle = 1$, since ${\sigma_{el}}/{\sigma_{tot}}=1/2$ and then
 starts  to decrease at the energies where this ratio goes beyond the black
  disk limit $1/2$.
Such qualitative
 nonmonotonous energy dependence of inelasticity is the result of transition to
  the antishadowing
 scattering regime. It is worth noting  that the maximum in inelasticity energy
 dependence is
 correlated with the minimum of the relative fraction of protons in the cosmic rays.

 It should be noted that the behaviour of the ratio
  ${\sigma_{el}}/{\sigma_{tot}}$ when it goes to unity at
  $s\to\infty$ does not mean decreasing energy dependence of $\sigma_{inel}$.
  The inelastic cross--section $\sigma_{inel}$ increases monotonically
  and it grows as $\ln s$ at $s\to\infty$. Therefore the depth of shower
  maximum which is related to the probability of inelastic interactions
  would become shallower with energy. Its energy dependence is not affected
  by the dominating role of elastic scattering which occurs first at the small
  impact parameters.

The relation of the knee and other features observed in the cosmic rays measurements
 with the
modification of particle generation mechanism is under discussion since the time
of their discoveries.
We would like to point out here one particular realization of such approach where
the corresponding generation mechanism is strongly affected by unitarity effects and
 the energy region between knee and ankle is related to
 the transition region to the antishadow scattering mode, i.e. the real energy spectrum
  $F_0(E)$
is modulated  by the significant variation of the scattering matrix $S$ in the energy region
which starts at $E_1\simeq 10^5$ GeV and ends at $E_2\simeq 10^9$ GeV and this
 results in the  regularities in the observed
spectrum $F(E)$. Below the energy $E_1$ and beyond the energy $E_2$ variation
of scattering matrix is slow and the primary energy spectrum $F_0$ is not affected.
 This hypothesis is not only but would be one of the natural explanations
of the observed cosmic rays regularities.


\small
\begin{thebibliography}{99}
\bibitem{der}
A. De R\'ujula, hep-ph/0412094, astro-ph/0411763.
\bibitem{engel}
R. Engel, T.K. Gaisser, P. Lipari, T. Stanev,
Phys. Rev. D 58 (1998) 014019.
\bibitem{crrev}
T. Stanev, astro-ph/0411113;\\  S.I. Nikolsky,  V.G. Sinitsina,
Phys. Atom. Nucl. 67 (2004) 1900, ;\\  A.A. Petrukhin,
Proc. of the 28th International Cosmic Ray Conferences (ICRC 2003), Tsukuba,
Japan, 31 Jul - 7 Aug 2003, 275.
\bibitem{her}
 J. R. H\"{o}randel, talk at
19th European Cosmic Ray Symposium, Florence, Italy, 30 Aug - 3 Sep 2004;
astro-ph/0501251.
\bibitem{ashd}
S. M. Troshin, N. E. Tyurin, Phys. Lett. B 316 (1993) 175.
\bibitem{echn}
S. M. Troshin, N. E. Tyurin,
Phys. Part. Nucl. 35 (2004) 555.
\bibitem{s0}
S. M. Troshin, N. E. Tyurin,
Eur. Phys. J. C  21 (2001) 679 ;\\
V. A. Petrov, A. V. Prokudin, S. M. Troshin, N. E. Tyurin,
J. Phys. G 27 (2001) 2225.
\bibitem{sc}
S. M. Troshin, Phys. Lett. B 597 (2004) 391.
\bibitem{gsp}
S. M. Troshin, N. E. Tyurin,
Eur. Phys. J. C  39 (2005) 435.
\bibitem{hern}
J. R. H\"{o}randel, J. Phys. G,  29 (2003) 2439.
\bibitem{shabel}
Yu. M. Shabelski, R. M. Weiner, G. Wilk, Z. Wlodarczyk, J. Phys. G, (1992) 1281.
\bibitem{dias}
J. Dias de Deus, Phys. Rev. D, 32 (1985) 2334 ;\\
S. Barshay, Y. Ciba, Phys. Lett. B 167 (1985) 449.
\end{thebibliography}
\end{document}